\documentclass[12pt,preprint]{aastex}

\shorttitle{Self-similar shocks: emergence into vacuum}

\begin{document}

\title{Self-similar solutions for relativistic shocks \\ emerging from stars with polytropic envelopes}

\author{Margaret Pan and Re'em Sari}

\affil{130-33 Caltech, Pasadena, CA 91125}

\email{mpan@astro.caltech.edu, sari@tapir.caltech.edu}

\begin{abstract}
We consider a strong ultrarelativistic shock moving through a star
whose envelope has a polytrope-like density profile. When the shock is
close to the star's outer boundary, its behavior follows the
self-similar solution given by Sari (2005) for implosions in planar
geometry. Here we outline this solution and find the asymptotic
solution as the shock reaches the star's edge. We then show that the
motion after the shock breaks out of the star is described by a
self-similar solution remarkably like the solution for the motion
inside the star. In particular, the characteristic Lorentz factor,
pressure, and density vary with time according to the same power laws
both before and after the shock breaks out of the star.  After
emergence from the star, however, the self-similar solution's
characteristic position corresponds to a point behind the leading edge
of the flow rather than at the shock front, and the relevant range of
values for the similarity variable changes. Our numerical integrations
agree well with the analytic results both before and after the
shock reaches the star's edge.
\end{abstract}

\keywords{hydrodynamics --- shock waves --- stars: general}

\section{Introduction}

The surge of activity over the past decade or so in the fields of
supernovae and of gamma-ray bursts and their afterglows has led to
renewed investigation into the behavior of strong shocks.  Much of the
analytic work on strong shock propagation to date has focused on
self-similar solutions to the hydrodynamic equations.  In these
solutions, the profiles of the hydrodynamic variables as functions of
position have constant overall shapes whose time evolution consists
simply of scalings in amplitude and position.  As a result,
self-similarity allows us to reduce the nominal system of
two-dimensional partial differential hydrodynamic equations to a
system of one-dimensional ordinary differential equations. The
existence of self-similar solutions thus enables a significant
simplification of problems free of spatial scales in regions far from
the initial conditions. The best-known such solutions are the
pioneering Sedov-Taylor solutions for non-relativistic point
explosions propagating into surroundings with power-law density
profiles \citep{sedov46,vonneumann47,taylor50}.

Self-similar solutions are traditionally divided into two categories
(see, for example, \cite{zeldovich67} for a detailed
discussion). `Type I' solutions are those in which the time evolution
of the shock position and hydrodynamic variables follows from global
conservation laws such as energy conservation. The Sedov-Taylor
solutions are Type I; their ultrarelativistic analogues were found by
\cite{blandford76}. By contrast, global conservation laws are useless
in `Type II' solutions, which are instead characterized by the
requirement that the solution remain well-behaved at a singular point
known as the `sonic point'. If, for instance, the density of the
surroundings falls off very quickly with distance, Type II solutions
found by \cite{waxman93} for non-relativistic spherical explosions
hold instead of the Sedov-Taylor solutions and relativistic solutions
found by \cite{best00} hold instead of the Blandford-McKee solutions.

Here we study the case of an ultrarelativistic shock wave moving
outwards through a star whose envelope has a polytrope-like density
profile. After the shock front reaches the outer edge of the star, an
event we refer to as `breakout', the shock front itself ceases to
exist but the shocked fluid continues outward into the vacuum
originally surrounding the star. We focus on the flow at times just
before and just after breakout. As explained in \S 2, the shock
evolution just inside the star's surface is identical to that expected
for an imploding planar shock in a medium with a power-law density
profile. Such a shock follows a Type II self-similar solution as
discussed by \cite{sari05} and \cite{nakayama05} and outlined briefly
here. \S 3 describes the asymptotic solution as the shock front
reaches the surface of the star, a singular point. In \S 4 we
investigate the flow after breakout.  We show that the self-similar
solution for the evolution inside the star also describes the behavior
outside the star except in that a different range of the similarity
variable applies and in that the physical interpretation of the
characteristic position changes. We show in \S 5 that the analytic
results of \S 2, 3, 4 agree with our numerical integrations of the
relativistic time-dependent hydrodynamic equations, and in \S 6 we
summarize our findings. Throughout our discussion, we take the
speed of light to be $c=1$.

\section{Shock propagation within the star}

Since we are interested in the shock after it has reached the envelope
or the outermost layers of a star, we assume that the mass and
distance lying between the shock front and the star's outer edge are
much less than the mass and distance between the shock front
and the star's center. In this region, we can take the star's gravity
$g$ to be constant and the geometry to be planar. We also assume that
the stellar envelope has a polytrope-like equation of state, that is,
$p\propto\rho^{q}$ where $p$ is the pressure, $\rho$ is the mass density,
and $q$ is a constant. This type of equation of state occurs
in various contexts including fully convective stellar envelopes, in
which case $q$ is the adiabatic index; radiative envelopes where the
opacity has a power-law dependence on the density and temperature; and
degenerate envelopes.

Under these assumptions we can find the density profile from hydrostatic
equilibrium and the equation of state as follows. Let $x$ be the radial
coordinate such that $x=0$ at the star's surface and $x<0$ inside the star.
Then
\begin{equation}
0 = \frac{dp}{dx}+\rho g
\end{equation}
and with the boundary condition $\rho =p=0$ at the edge of the star,
we have
\begin{equation}
\frac{q}{q-1}\rho ^{q-1} \propto -gx
\end{equation}
\begin{equation}
\rho \propto (-x)^{1/(q-1)} = (-x)^{-k} \;\;\; .
\end{equation}
For convective and degenerate envelopes, $q$ is between $4/3$ and
$5/3$; for radiative envelopes with Kramers opacity, $q=30/17$. 
These give $k$ values between $-1$ and $-3$.

With the power-law density profile $\rho\propto(-x)^{-k}$, the
evolution of an ultrarelativistic shock propagating through the
envelope is given by a Type II converging planar self-similar solution
to the hydrodynamic equations representing energy, momentum, and mass
conservation,
\begin{equation}
\label{e_conserve}
\frac{\partial}{\partial t}\left[\gamma^2(e+\beta^2 p)\right]
+ \frac{\partial}{\partial x}\left[\gamma^2\beta(e+p)\right]
 = 0
\end{equation}
\begin{equation}
\label{p_conserve}
\frac{\partial}{\partial t}\left[\gamma^2\beta(e+p)\right]
+ \frac{\partial}{\partial x}\left[\gamma^2(\beta^2 e+p)\right]
 = 0
\end{equation}
\begin{equation}
\label{n_conserve}
\frac{\partial}{\partial t}(\gamma n)
+ \frac{\partial}{\partial x}(\gamma\beta n)
 = 0 \;\;\; ,
\end{equation}
with the ultrarelativistic equation of state
\begin{equation}
p = \frac{1}{3}e \;\;\; .
\end{equation} 
Here we will simply state the solution; for a detailed derivation see
\cite{sari05} or \cite{nakayama05}. We assume the effect of the star's
gravity on the shock propagation is negligible. Following
\cite{sari05}, we let $R(t)$ be the solution's characteristic
position, which we choose to be the position of the shock front while
the shock is within the star. We take $t=0$ at the time the shock
reaches the star's surface ($R=0$), and we take $R<0$ when $t<0$. We
take $\Gamma(t)$, $P(t)$, and $N(t)$ to be respectively the
characteristic Lorentz factor, pressure, and number density, and we
define
\begin{equation}
\label{char_powerlaws}
\frac{t\dot{\Gamma}}{\Gamma} = -\frac{m}{2} \;\;\; , \;\;\;
\frac{t\dot{P}}{P} = -m-k \;\;\; , \;\;\;
\frac{t\dot{N}}{N} = -\frac{m}{2} - k \;\;\; .
\end{equation}
Following \cite{blandford76}, we define the similarity variable as
\begin{equation}
\label{chi}
\chi = 1 + 2(m+1)\frac{R-x}{R/\Gamma^2} \;\;\; .
\end{equation}
Note that for $R<0$, $x\leq R$ and the relevant range in $\chi$ is
$-\infty<\chi<1$ as long as $m>-1$. We define the hydrodynamic
variables---the Lorentz factor $\gamma$, the pressure $p$, and the
number density $n$---as follows:
\begin{equation}
\label{gammadef}
\gamma^2(x,t) = \frac{1}{2}\Gamma^2(t)g(\chi)
\end{equation}
\begin{equation}
\label{pdef}
p(x,t) = P(t)f(\chi)
\end{equation}
\begin{equation}
\label{ndef}
n(x,t) = N(t)\frac{h(\chi)}{g^{1/2}(\chi)} \;\;\; .
\end{equation}
Here $g$, $f$, and $h$ give the profiles of $\gamma$, $p$, and $n$;
expressions for the dependence of $m$ on $k$ and for $g$, $f$, $h$ as
functions of $\chi$ make up the entire self-similar solution. The
above definitions and the ultrarelativistic hydrodynamic equations in
planar geometry put the sonic point, the point separating fluid
elements which can communicate with the shock front via sound waves
from those which cannot, at $g\chi=4-2\sqrt{3}$. Requiring that the
solution pass smoothly through this point gives
\begin{equation}
\label{m}
m = \left(3-2\sqrt{3}\right)k
\end{equation}
\begin{equation}
\label{g_general}
g = C_g\left| \frac{g\chi}{3k-2k\sqrt{3}+1}-2(2+\sqrt{3})\right|
            ^{-\left( 3-2\sqrt{3}\right)k}
\end{equation}
\begin{equation}
\label{f_general}
f = C_f\left|-g\chi-2k\sqrt{3}+4+2\sqrt{3}\right|^{-\left(4-2\sqrt{3}\right)k}
\end{equation}
\begin{equation}
\label{h_general}
h = C_h
    \left|g\chi +2k\sqrt{3}-4-2\sqrt{3}\right|
         ^{-\frac{\left( 2\sqrt{3}-3\right)(2k-1)k}{(-1+k\sqrt{3}-\sqrt{3})}}
    |g\chi -2|^{\frac{k}{-1+k\sqrt{3}-\sqrt{3}}} \;\;\; .
\end{equation}
The boundary conditions $g(\chi=1)=f(\chi=1)=h(\chi=1)=1$ which hold
inside the star allow us to determine the constants of integration
$C_g$, $C_f$, $C_h$ and write
\begin{equation}
\label{g_inside}
g = \left[ \frac{-g\chi - 2k\sqrt{3} + 4 + 2\sqrt{3}}
                {-1 - 2k\sqrt{3} + 4 + 2\sqrt{3}}
    \right] ^{-\left(3-2\sqrt{3}\right)k}
\end{equation}
\begin{equation}
\label{f_inside}
f = \left[ \frac{-g\chi - 2k\sqrt{3} + 4 + 2\sqrt{3}}
                {-1 - 2k\sqrt{3}+4+2\sqrt{3}}
    \right]^{-(4-2\sqrt{3})k}
\end{equation}
\begin{equation}
\label{h_inside}
h = \left[ \frac{g\chi +2k\sqrt{3}-4-2\sqrt{3}}
                {1+2k\sqrt{3}-4-2\sqrt{3}}
    \right]^{-\frac{\left( 2\sqrt{3}-3\right)(2k-1)k}
                   {-1+k\sqrt{3}-\sqrt{3}}}
    \left[ 2-g\chi \right]^{\frac{k}{-1+k\sqrt{3}-\sqrt{3}}} \;\;\; .
\end{equation}

\section{Transition at breakout}

To know what happens to the shocked material after the shock front
emerges from the star, we need the behavior of the shock just as the
front reaches the surface---the `initial conditions' for the evolution
of the shock after breakout. Specifically, we are interested in the
limiting behavior of each fluid element and in the asymptotic profiles
of $\gamma$, $p$, and $n$ as functions of $x$ as $t$ and $R$ approach
0. 

The limiting behavior of a given fluid element may be found as
follows. Due to the self-similarity, we know the time taken for
$\gamma$, $p$, and $n$ of a given fluid element to change
significantly is the timescale on which $R$ changes by an amount of
order itself.  Since $R$ can change by this much only once between the
time a given fluid element is shocked and the time the shock breaks
out of the star, the limiting values of $\gamma$, $p$, and $n$ for
that fluid element should be larger only by a factor of order unity
from their values when the fluid element was first shocked.

We can also find the scalings of $\gamma$, $p$, and $n$ with $x$ at
breakout via simple physical arguments. We denote by $x_{0}$,
$\gamma_{0}$, $p_0$, $n_0$ the position, Lorentz factor, pressure, and
number density of a fluid element just after being shocked and by
$x_{f}$, $\gamma_f$, $p_f$, $n_f$ those values at the time the shock
breaks out. Since the shock accelerates to infinite Lorentz factors,
and since, as we found above, the Lorentz factor of a given fluid
element remains constant up to a numerical factor, this fluid element
will lag behind the shock by $x_{f}\sim x_{0}/\gamma _{0}^{2}$ at
$t=0$. Eq.~\ref{char_powerlaws} gives $\Gamma\sim t^{-m/2}$, so we
have $\gamma _{0}\sim (-x_{0})^{-m/2}$; then $\gamma_f\sim
((-x_f)\gamma_f^2)^{-m/2}$ or 
\begin{equation}
\gamma_f\sim(-x_f)^{-m/2(m+1)} \;\;\; .
\end{equation}
Likewise, since $P\sim t^{-m-k}$ and $N\sim t^{-m/2-k}$, we have
$p_0\sim x_0^{-m-k}$ and $n_0\sim x_0^{-m/2-k}$; then
\begin{equation}
p_f\sim ((-x_f)\gamma_f^2)^{-m-k}\sim (-x_f)^{-(m+k)/(m+1)}
\end{equation}
\begin{equation}
n_f\sim (-x_f)^{-(m/2+k)/(m+1)} \;\;\; .
\end{equation}

We can use the equations for the solution before breakout to perform
equivalent calculations of the limiting behavior of fluid elements and
asymptotic profiles of $\gamma$, $p$, $n$. For the limiting behavior
of a fluid element, we take the advective time derivative of $g\chi$
and use the result to relate $\gamma$ and $g$ to time for that fluid
element. The advective derivative is given by
\begin{equation}
\label{ddt_advective}
\frac{D}{Dt}
 = \frac{\partial }{\partial t} + \beta \frac{\partial }{\partial r}
 = \dot{\Gamma}\frac{\partial }{\partial \Gamma }
   + \dot{P}\frac{\partial}{\partial P}
   + \frac{m+1}{t}(2/g-\chi) \frac{\partial}{\partial \chi}
\;\;\; .
\end{equation}
We apply this derivative to Eq.~\ref{g_inside} to get
\begin{equation}
\frac{D(g\chi)}{D\log t}
 = \frac{(2-g\chi) \left( g\chi-4-2\sqrt{3}+2\sqrt{3}k\right)}
        {\left( g\chi-4-2\sqrt{3}\right) }
\end{equation}
and integrate to get
\begin{equation}
\label{t_vs_gchi}
t/t_{0}
 = \left| g\chi -2\right| ^{(3+\sqrt{3})/(3k-\sqrt{3}-3)}
   \left|\frac{g\chi -4-2\sqrt{3}+2k\sqrt{3}}{1-4-2\sqrt{3}+2k\sqrt{3}}\right|
        ^{-3k/\left(-\sqrt{3}-3+3k\right) }
\end{equation}
where $t_0$ is the time at which the fluid element is shocked, that
is, when $g=\chi=1$. When $\left| g\chi \right| \gg 1$---which becomes
true everywhere behind the shock front as $t\rightarrow 0$---this
simplifies to
\begin{equation}
t/t_{0}
\label{t_vs_gchi_limit}
 \simeq \left|g\chi \right| ^{-1}
        \left|1-4-2\sqrt{3}+2k\sqrt{3}\right|^{3k/\left(-\sqrt{3}-3+3k\right)}
\end{equation}
and Eq.~\ref{g_inside} simplifies to
\begin{equation}
\label{g_inside_limit}
g \simeq \left[ \frac{-g\chi }{-1-2k\sqrt{3}+4+2\sqrt{3}}\right]
              ^{-\left( 3-2\sqrt{3}\right) k} \;\;\; .
\end{equation}
We substitute Eq.~\ref{t_vs_gchi_limit} into Eq.~\ref{g_inside_limit} to
get the limiting Lorentz factor of the fluid element as $t\rightarrow 0$:
\begin{equation}
\gamma = \gamma _{0}
         \left| 1-4-2\sqrt{3}+2k\sqrt{3}\right|
         ^{-\left(3-3\sqrt{3}\right) k 
           / \left(2\left(-\sqrt{3}-3+3k\right)\right)}
\end{equation}
which is greater only by a numerical factor than the initial Lorentz
factor $\gamma _{0}$ that the fluid element received right after being
shocked. To relate the limiting $p$, $n$ to $p_0$, $n_0$, we likewise
take Eqs.~\ref{f_inside}, \ref{h_inside} in the limit $|g\chi|\gg 1$ and
use Eqs.~\ref{t_vs_gchi_limit}, \ref{g_inside_limit} with the results to get
\begin{equation}
p = p_0 \left| 1-4-2\sqrt{3}+2k\sqrt{3}\right|
             ^{-\left( 6-2\sqrt{3}\right) k/\left( -\sqrt{3}-3+3k\right) }
\end{equation}
\begin{equation}
n = n_0 \left| 1-4-2\sqrt{3}+2k\sqrt{3}\right|
             ^{-\frac{\left(4k+k\sqrt{3}-3-\sqrt{3}\right)
                      \left( 3-2\sqrt{3}\right) k}
                     {2\left(k\sqrt{3}-1-\sqrt{3}\right) }}
\end{equation}
which again differ only by numerical factors from their values just
after the fluid element is shocked.  This is consistent with the
behavior given above by simple physical considerations.

For the analogous calculation of the asymptotic profiles of $\gamma$,
$p$, and $n$, we cannot simply apply Eqs.~\ref{gammadef}, \ref{pdef},
\ref{ndef}: Eqs.~\ref{char_powerlaws}, \ref{chi} require that
$\chi\rightarrow -\infty$ everywhere behind the shock and $\Gamma$,
$P$, and $N$ diverge as $t\rightarrow 0$. Instead we take the
$t\rightarrow 0$ limit at a fixed position $x$. First we have
\begin{equation}
\chi = 1+2(m+1)(1-x/R)\Gamma ^{2}
     \simeq 2(m+1)(-x/R)\Gamma ^{2} \;\;\; .
\end{equation}
With Eqs.~\ref{gammadef}, \ref{g_inside_limit} this gives
\begin{equation}
2\gamma ^{2}/\Gamma ^{2} = g
 = \left[\frac{-4(m+1)(-x)\gamma^{2}/R}{-1-2k\sqrt{3}+4+2\sqrt{3}}
   \right]^{-m}
\end{equation}
and
\begin{equation}
\label{gamma_inside_limit}
\gamma
 = \left[2\frac{(-R)^{-m}}{\Gamma ^{2}}\right] ^{-1/2(1+m)}
   \left[ \frac{4(m+1)}{-1-2k\sqrt{3}+4+2\sqrt{3}}\right]^{-m/2(1+m)}
   (-x)^{-m/2(1+m)} \;\;\; .
\end{equation}
This is consistent with our qualitative discussion; the coefficient in
the qualitative relation is a numerical factor times the constant
$(-R)^{-m}/\Gamma ^{2}$. For the $p$ and $n$ profiles, we apply a
similar analysis to the expressions for $f$ and $h$ in the limit
$t\rightarrow 0$.
\begin{equation}
\label{p_inside_limit}
p = P\left[2\frac{(-R)}{\Gamma^2}\right]^\frac{m+k}{1+m}
    \left[\frac{4(m+1)}{-1-2k\sqrt{3}+4+2\sqrt{3}}\right]^{-\frac{m+k}{1+m}}
    (-x)^{-\frac{m+k}{1+m}}
\end{equation}
\begin{equation}
\label{n_inside_limit}
n = N\left[2\frac{(-R)}{\Gamma^2}\right]^\frac{m/2+k}{1+m}
    \left[4(m+1)\right]^{-\frac{m/2+k}{1+m}}
    \left[-1-2k\sqrt{3}+4+2\sqrt{3}\right]
         ^{\frac{m/2+k}{1+m} + \frac{k}{-1+k\sqrt{3}-\sqrt{3}}}
    (-x)^{-\frac{m/2+k}{1+m}}
\end{equation}
These results are likewise consistent with our qualitative discussion.

\section{Evolution after breakout}

\subsection{Self-similar solution}

Since the breakout itself does not introduce new spatial scales into
the flow, we expect the motion after breakout to remain
self-similar. However, as the shock Lorentz factor diverges at $t=0$,
we cannot continue to associate the characteristic position, Lorentz
factor, pressure, and number density with the values at the shock
front after breakout. So we begin by providing physical motivation for
a different characteristic Lorentz factor and exploring the
implications of this choice.

We note that after breakout each fluid element expands and accelerates
over time until the element's internal energy has been converted
entirely into bulk motion.  Given a relativistic strong shock, the
internal energy of a shocked fluid element in the frame moving with
the fluid is comparable to the bulk kinetic energy of the fluid
element. This implies that the fluid element's final bulk Lorentz
factor should be much greater than the value of the shock Lorentz
factor just after the fluid element was shocked. The timescale $t_x$
for the resulting expansion and acceleration is the time over which
the fluid element's size and Lorentz factor change by a factor of
order unity. For a fluid element located at $-x$ and with Lorentz
factor $\gamma_x$ at $t=0$, the time of breakout, this timescale is
$t_x=x\gamma_x^2$ due to relativistic beaming. That every time $t>0$
is thus associated in a scale-independent way with a particular $t_x$
and $\gamma$ suggests that we pick $\Gamma(t=t_x)=\gamma_x$ to be the
characteristic Lorentz factor.

To see how $\Gamma$ evolves with time, we use
$\gamma\propto(-x)^{-m/2(1+m)}$ from Eq.~\ref{gamma_inside_limit} with
the $t_x$ relation above to get $\Gamma\propto t^{-m/2}$. For the
characteristic pressure $P$ and number density $N$,
Eqs.~\ref{p_inside_limit}, \ref{n_inside_limit} likewise give
$P\propto t^{-m-k}$ and $N\propto t^{-m/2-k}$. In other words,
Eq.~\ref{char_powerlaws} holds after breakout with exactly the same
$k$, $m$ that apply inside the star. The characteristic position $R$
is again the position which evolves according to the Lorentz factor
$\Gamma$: $\dot{R}\simeq 1-1/2\Gamma^2$. Since the hydrodynamic
equations still hold as well, Eqs.~\ref{chi}, \ref{g_general},
\ref{f_general}, \ref{h_general} must remain valid when $t>0$.

To find the complete solution after breakout we need to specify the
boundary conditions. We proceed by looking at the behavior of the
similarity variables $\chi$, $g$, $f$, $h$. The relevant range in
$\chi$ depends on $R$, and while the relation between $R$ and $\Gamma$
is the same before and after breakout, $R$ after breakout is not the
position of the shock front. Instead, the front has infinite Lorentz
factor and $R$ lags further and further behind the shock with
increasing time. A nice physical interpretation exists for $R$ after
breakout. $R$ tracks the position corresponding to a fluid element
which has expanded by a factor of order unity, so $R$ marks the
transition in position between fluid elements which have expanded and
accelerated significantly since being shocked and fluid elements whose
size and speed have remained roughly constant. Since it takes longer
for fluid elements with smaller Lorentz factors to expand and
accelerate significantly, $R$ moves backward relative to the leading
edge of the flow at $x=t$. Because $R$ becomes positive after
breakout, the range of possible $x$ in the solution outside the star
is $x\leq t$. Then $\chi=0$ at the `front' $x=t$, and the relevant
range in $\chi$ in the solution outside is $0<\chi<\infty$ rather than
$-\infty<\chi<1$.

\begin{figure}
\centerline{\hbox{\plotone{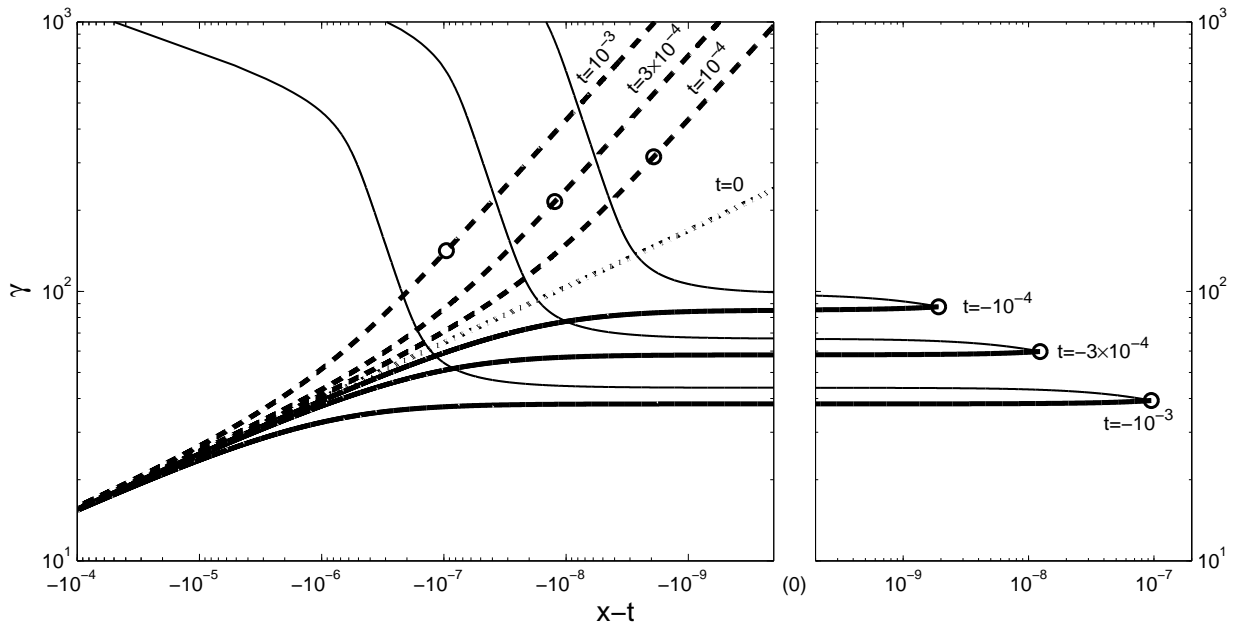}}}
\caption{Profiles of $\gamma$ as a function of position (heavy lines)
at seven different times marked on the figure and trajectories of
three fluid elements in position-Lorentz factor space (thin
lines). Fluid elements at the characteristic positions $R$ are
marked by open circles. We use $x-t$ as the position coordinate to
allow easy comparison of the profiles. 
The $t=0$ curve (heavy dotted line) is the asymptotic profile
corresponding to the pure power law $\gamma\propto (-x)^{-m/2(1+m)}$
given in Eq.~\ref{gamma_inside_limit}. The profiles with $t<0$ (heavy
solid lines) are given by Eqs.~\ref{gammadef}, \ref{g_inside} and the
profiles with $t>0$ (heavy dashed lines) are given by
Eqs.~\ref{gammadef}, \ref{g_outside}.  
When $t<0$, the natural choices for $R$ and $\Gamma$ are respectively
the location of the shock front and the Lorentz factor of the front.
When $t>0$, a fluid element at position $R$ has accelerated by a
factor of order unity and its Lorentz factor is of order $\Gamma$.  So
the positions $R$ lie just above the `knees' in the profiles, which
separate fluid elements which have already expanded from those which
have not. When $|x-t|\gg R/\Gamma^2$ or, equivalently, $|\chi|
\rightarrow \infty$, all profiles approach the $t=0$ power law since
at $t=0$, $|\chi|\rightarrow\infty$ everywhere behind the front.  When
$|x-t|\ll R/\Gamma^2$, the $t<0$ profiles approach a constant
($\gamma\rightarrow\Gamma/\sqrt{2}$) and the $t>0$ profiles approach
$\gamma\propto |x-t|^{-1}$ ($g\propto\chi^{-1}$ from
Eq.~\ref{g_outside}). Because every fluid element is always accelerating,
the $t<0$ profiles always lie below the $t=0$ power law and the $t>0$
profiles are always above the $t=0$ power law.
Trajectories of individual fluid elements before breakout are given by
Eq.~\ref{t_vs_gchi}. After breakout, Eq.~\ref{t_vs_gchi} still
applies.  The power laws relating $t$ to $g\chi$ stay the same after
breakout since the equations for $g$ before and after breakout are
nearly identical; also, matching the pre- and post-breakout
trajectories at $t=0$ gives the same $|t_0|$ in the evolution both
before and after $t=0$.}
\label{gprofilevst}
\end{figure}

Far behind $x=t$, the profiles of $\gamma$, $p$, and $n$ before
breakout must coincide with the profiles after breakout. We know this
because at a given time after breakout, sound waves carrying the
information that breakout occurred can only have traveled a finite
distance behind the shock front; material further behind continues to
flow as if the breakout had never occurred. Also, the two sets of
profiles must coincide at $t=0$, when everything is far behind the
front.  To phrase this requirement on the profiles in terms of the
similarity variable, $g(\chi\rightarrow-\infty)$,
$f(\chi\rightarrow-\infty)$, and $h(\chi\rightarrow-\infty)$ before
breakout must coincide with $g(\chi\rightarrow\infty)$,
$f(\chi\rightarrow\infty)$, and $h(\chi\rightarrow\infty)$ after
breakout. Then as $\chi\rightarrow \infty$ after breakout,
$g,f,h\rightarrow 0$ and $g\chi\rightarrow\infty$. In addition, the
constants $C_g$, $C_f$, $C_h$ in Eqs.~\ref{g_general},
\ref{f_general}, \ref{h_general} must be the same for both the pre-
and post-breakout solutions. In other words, the solutions before and
after breakout, as specified by Eqs.~\ref{chi}, \ref{g_general},
\ref{f_general}, \ref{h_general} and expressions for $C_g$, $C_f$,
$C_h$, are the same; only the relevant ranges in $\chi$ and the
physical interpretations of the variables differ. So the expressions
for $g$, $f$, $h$ after breakout are
\begin{equation}
\label{g_outside}
g = \left[ \frac{g\chi + 2k\sqrt{3} - 4 - 2\sqrt{3}}
                {-1 - 2k\sqrt{3} + 4 + 2\sqrt{3}}
    \right] ^{-\left(3-2\sqrt{3}\right)k}
\end{equation}
\begin{equation}
\label{f_outside}
f = \left[ \frac{g\chi + 2k\sqrt{3} - 4 - 2\sqrt{3}}
                {-1 - 2k\sqrt{3}+4+2\sqrt{3}}
    \right]^{-(4-2\sqrt{3})k}
\end{equation}
\begin{equation}
\label{h_outside}
h = \left[ \frac{-g\chi -2k\sqrt{3}+4+2\sqrt{3}}
                {1+2k\sqrt{3}-4-2\sqrt{3}}
    \right]^{-\frac{\left( 2\sqrt{3}-3\right)(2k-1)k}
                   {-1+k\sqrt{3}-\sqrt{3}}}
    \left[ g\chi-2 \right]^{\frac{k}{-1+k\sqrt{3}-\sqrt{3}}} \;\;\; .
\end{equation}
The boundary conditions after breakout are given explicitly by $g = f
= 1$ and
$h=\left(5+4\sqrt{3}-4\sqrt{3}k\right)^{k/\left(-1-\sqrt{3}+k\sqrt{3}\right)}$
at $\chi = 7+4\sqrt{3}-4\sqrt{3}k$. A graphical comparison between the
pre- and post-breakout $\gamma$ vs. position profiles is given in
Figure~\ref{gprofilevst} along with sample trajectories of fluid
elements.

\subsection{Type I or Type II?}

While the flow before breakout follows a Type II self-similar
solution, the solution describing the flow after breakout contains
elements of Type I and Type II solutions. Unlike the Type II solution
which applies before breakout, the post-breakout solution does not
contain a sonic point. Differentiating Eq.~\ref{g_outside} with
respect to $g\chi$ shows that the only local extremum of $g\chi$
occurs at $g=\infty$ or $\chi=0$, where
$g\chi=4+2\sqrt{3}-2k\sqrt{3}$; since $g\chi\rightarrow\infty$ as
$\chi\rightarrow\infty$, $g\chi$ must attain its global minimum at
$\chi=0$. But then for $k<0$ neither the sonic point,
$g\chi=4-2\sqrt{3}$, nor the other singular points, $g\chi=2$ and
$g\chi=4+2\sqrt{3}$, is included in the solution after breakout. A
more physical argument for the exclusion of the sonic point from the
post-breakout solution is that since each fluid element is
accelerating while $\Gamma$ decreases with time, the fluid element
moves forward relative to $R$ and its $\chi$ must decrease with
time. Using Eq.~\ref{ddt_advective} we see that $D\chi/Dt<0$ requires
$g\chi>2>4-2\sqrt{3}$ for every fluid element. Then the entire
post-breakout solution is causally connected as would be expected if
it were Type I.

Unlike Type I solutions, however, the solution after breakout contains
infinite energy: it can be thought of as representing a flow into
which a source at $x=-\infty$ feeds energy at a constant rate,
sustaining the acceleration of fluid elements further and further
behind the shock. As a result, global conservation laws do not apply
just as would be expected in a Type II solution. So the post-breakout
solution lies between the standard Type I and Type II solution
categories. While this unusual situation implies that, in principle,
the infinite energy contained in the solution can communicate with and
affect the region near $R$, the regions of the solution containing
this infinite energy lie arbitrarily far behind the front at $x=t$ and
therefore take arbitrarily long to communicate with the fluid near the
front. Similarly, in any application of the post-breakout solution,
the flow will be truncated at some position well behind $R$,
potentially introducing a spatial scale into the problem. However, the
solution is valid until information from the truncation region
propagates to areas close to $R$. The further the truncation from $R$,
the longer this will take.

\subsection{Relation to previous work}

The first analytic investigation of an ultrarelativistic planar shock
wave was performed by \cite{johnson71}. The problem they consider is
broadly similar to the one we discuss here, but our work differs in
important respects from theirs. First, \cite{johnson71} used the
method of characteristics in their work: they analyzed the flow
associated with the shock by tracing the paths of sound waves
travelling through the fluid. Our analysis uses the self-similarity of
the flow instead. So while some of their work can be applied to flows
moving through fluids with arbitrary decreasing density profiles,
their methods do not give profiles for the hydrodynamic variables as
functions of $x$ at a given time. By contrast, our self-similar
solutions require a power-law density profile inside the star but give
explicit profiles for the hydrodynamic variables. Second, the methods
used by \cite{johnson71} require initial conditions consisting of a
uniform stationary hot fluid about to expand into cold surroundings.
In our scenario the hot expanding fluid is never uniform or stationary
and always follows the self-similar profile specified by our solution.
The self-similarity analysis tells us that the solution is Type II, at
least before breakout; this implies that the asymptotic solution is
independent of the initial engine.

We can check that our asymptotic solution is consistent with the
findings of \cite{johnson71} by looking at the Lorentz factors of
individual fluid elements at very late times.  While in our
self-similar solution the fluid elements formally accelerate forever,
each fluid element must in practice stop accelerating when all of its
internal energy has been converted to bulk kinetic energy, or when
$p/n\sim\gamma f/h\sim 1$.  So we can estimate the final Lorentz
factor of a given fluid element from Eqs.~\ref{g_outside},
\ref{f_outside}, \ref{h_outside}. By taking the advective time
derivatives of $\gamma$ and of $f/h$ we can write differential
equations for their time evolution following a single fluid
element. These are
\begin{eqnarray}
\frac{D\gamma}{Dt}&
 =& \frac{\gamma}{t} \frac{\left(\sqrt{3}-3\right)k}{g\chi-4-2\sqrt{3}}
 \;\;\; \simeq \;\;\; \frac{\gamma}{t} \left(\frac{\sqrt{3}-1}{2}\right) \\
\frac{D(f/h)}{Dt}&
 =& \frac{(f/h)}{t}
    \left[\frac{(2-g\chi)+\left(g\chi-4-2\sqrt{3}+2k\sqrt{3}\right)}
               {g\chi-4-2\sqrt{3}}\right]
    \frac{k}{-1+k\sqrt{3}-\sqrt{3}} \\
&\simeq& \frac{(f/h)}{t} \left(\frac{-1}{\sqrt{3}}\right) \;\;\; .
\end{eqnarray}
In the last steps we have taken the limit of late times when the
accelerating fluid element approaches the shock front at $\chi=0$. In
this limit Eq.~\ref{g_outside} implies $g\rightarrow\infty$ and
$g\chi\rightarrow 4+2\sqrt{3}-2k\sqrt{3}$.  Let $\gamma_0$, $f_0$, and
$h_0$ be the values of the functions in question just after our fluid
element is shocked; then at late times $\gamma\gg \gamma_0$ so
$(f/h)/(f_0/h_0)\sim \gamma^{-1}$.  Integrating the above differential
equations then gives
\begin{equation}
\label{gamma_coldlimit}
\frac{\gamma}{\gamma_0}
 = \left(\frac{t}{t_0}\right)^{\left(\sqrt{3}-1\right)/2}
 \sim \gamma^{\left(3-\sqrt{3}\right)/2}
\;\;\; \longrightarrow \;\;\; 
\gamma\sim\gamma_0^{1+\sqrt{3}} \;\;\; .
\end{equation}
This agrees with the results of \cite{johnson71} for the final Lorentz
factor of the fluid in a strong ultrarelativistic shock propagating
into a cold medium with decreasing density. The agreement provides
additional support for our claim that the solution outside the star
behaves like the solution describing a standard planar shock up to the
initial conditions and the interpretation of the characteristic values
$R$, $\Gamma$, $P$, $N$. Note that the differences between the initial
conditions used in their work and in ours are unimportant to the
scaling law relating the final and initial Lorentz factors of a given
fluid element. This result agrees with the findings of \cite{tan01}
concerning the scaling law: partly because of uncertainty over the
different initial conditions, they used numerical simulations to check
the $\gamma\sim\gamma_0^{1+\sqrt{3}}$ result.


Recently, \cite{nakayama05} also investigated the problem of an
ultrarelativistic planar shock. While the self-similar solution they
give for the flow before breakout is identical to the one in
\cite{sari05} and outlined here, they do not give analytic results for
or a physical interpretation of the self-similar solution after
breakout.

\section{Comparison with numerical integrations}

\begin{figure}
\centerline{\hbox{\plotone{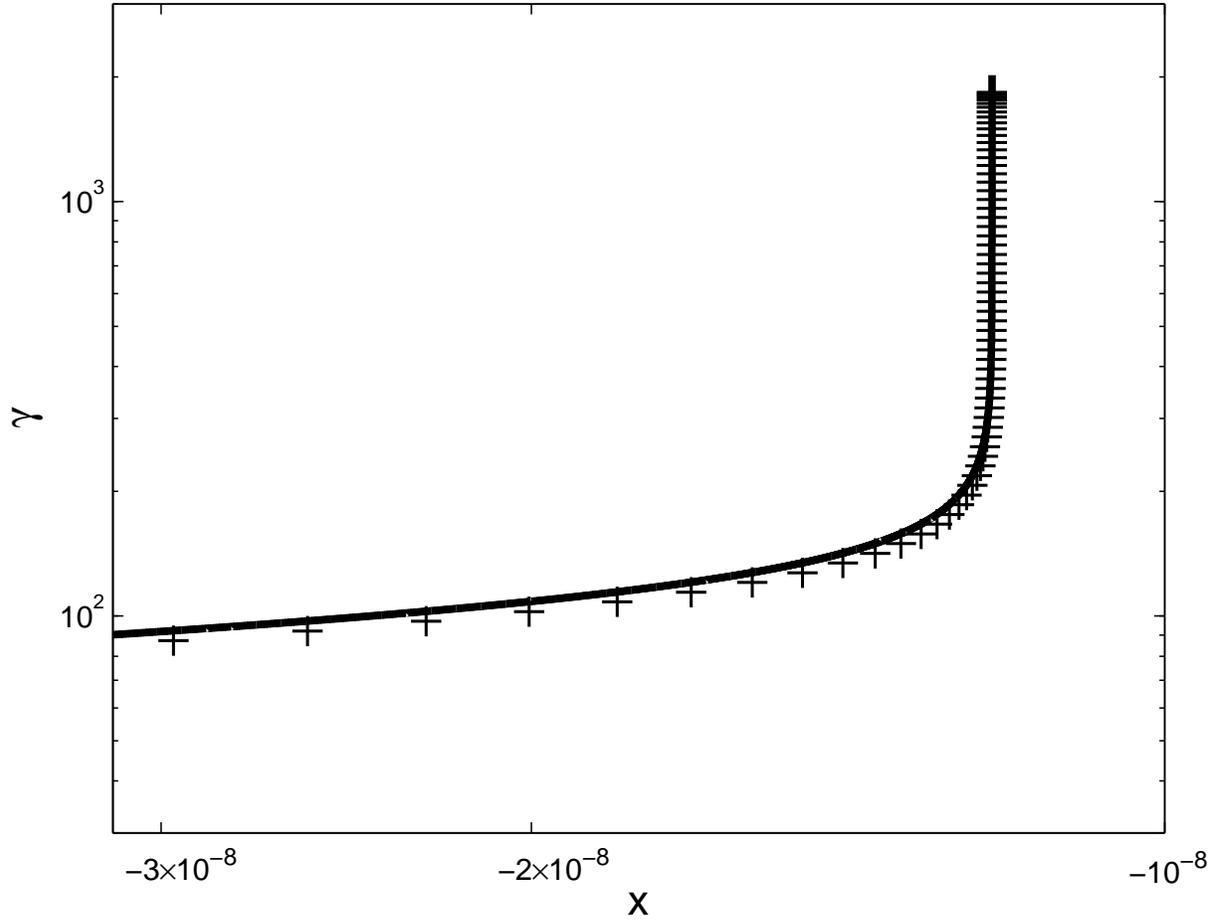}}}
\caption{Lorentz factor $\gamma$ as a function of position $x$ shortly
before the shock breaks out of the star. The density profile has
power-law index $k=-1.5$. The analytic profile taken from the
self-similar solution (solid line) agrees well with the numerical
profile (crosses).}
\label{gprofile_in}
\end{figure}

\begin{figure}
\centerline{\hbox{\plotone{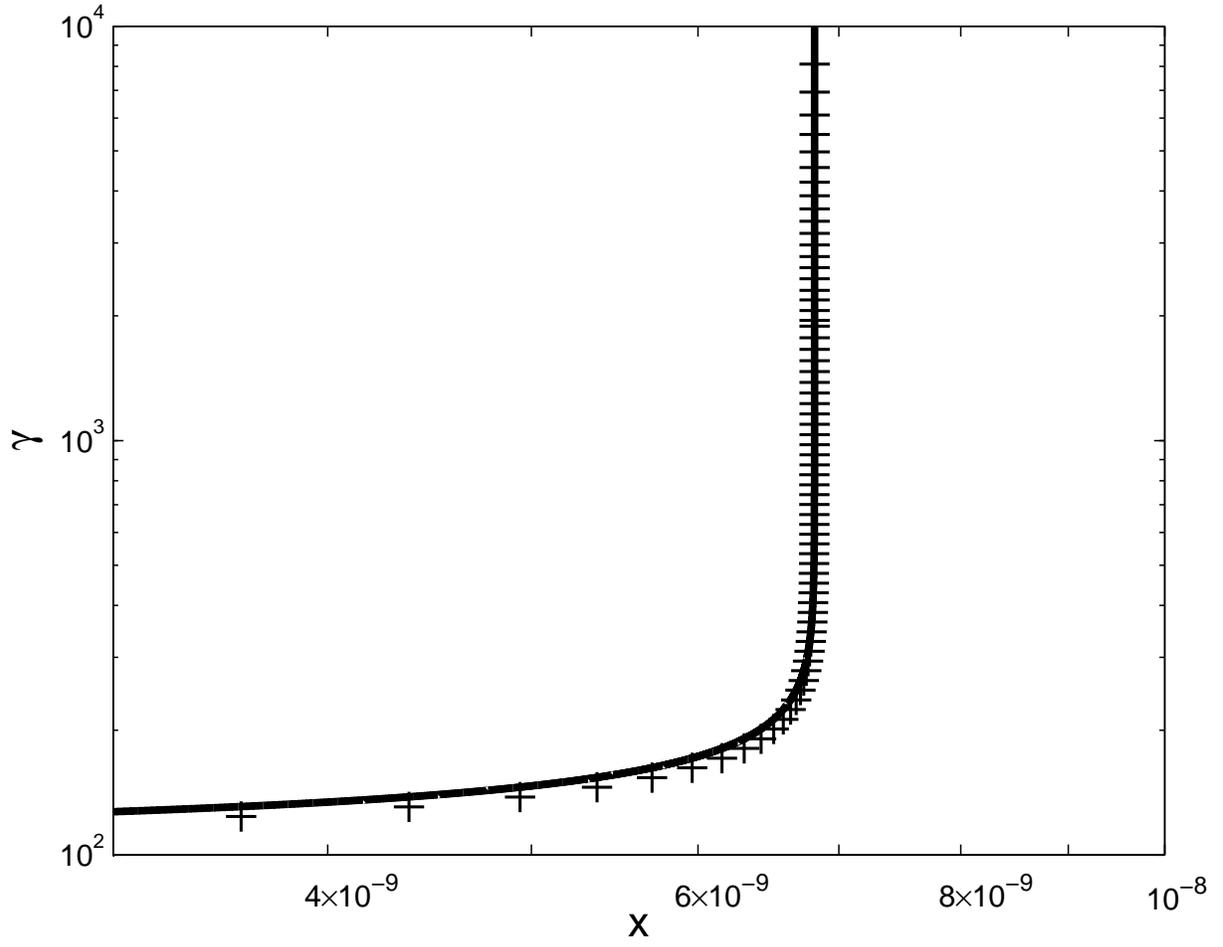}}}
\caption{Same as Figure~\ref{gprofile_in} but for a time shortly
after the shock emerges from the star.}
\label{gprofile_out}
\end{figure}

\begin{figure}
\epsscale{.9}
\centerline{\hbox{\plotone{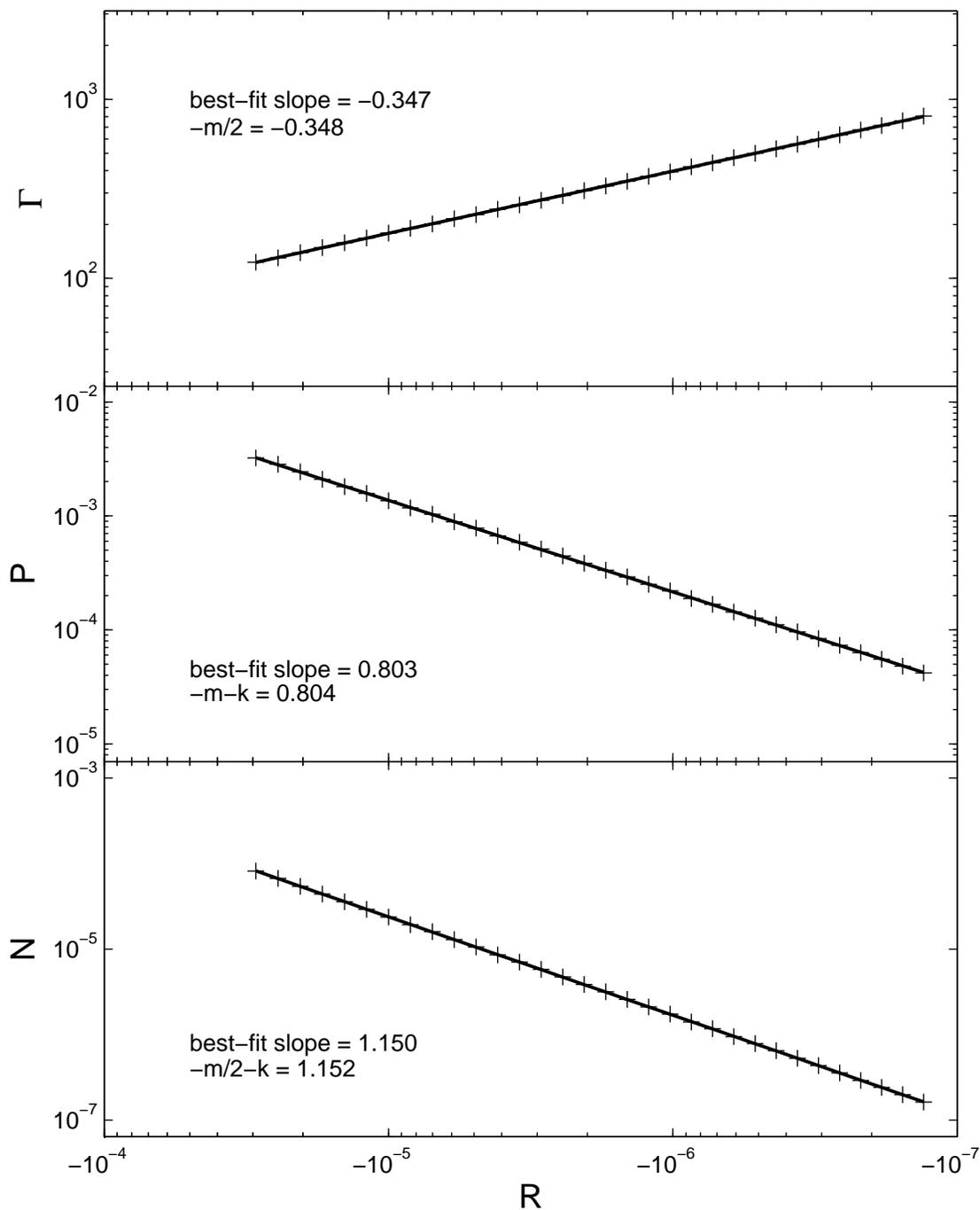}}}
\caption{Evolution of $\Gamma$ (top panel), $P$ (middle panel), and
$N$ (bottom panel) with $R$ while the shock is still inside the star.
The density profile has power-law index $k=-1.5$. The evolution of
$\Gamma$, $P$, $N$ with $R$ is equivalent to time evolution when
$\Gamma\gg 1$.  Crosses represent numerical data; solid lines are the
best-fit lines to the data. That the data are well fit by lines
implies that $\Gamma$, $P$, and $N$ do indeed evolve as power laws;
that the numerical and analytic slopes agree confirms that the
evolution is as expected.}
\label{gpnvstime_in}
\end{figure}

\begin{figure}
\epsscale{.9}
\centerline{\hbox{\plotone{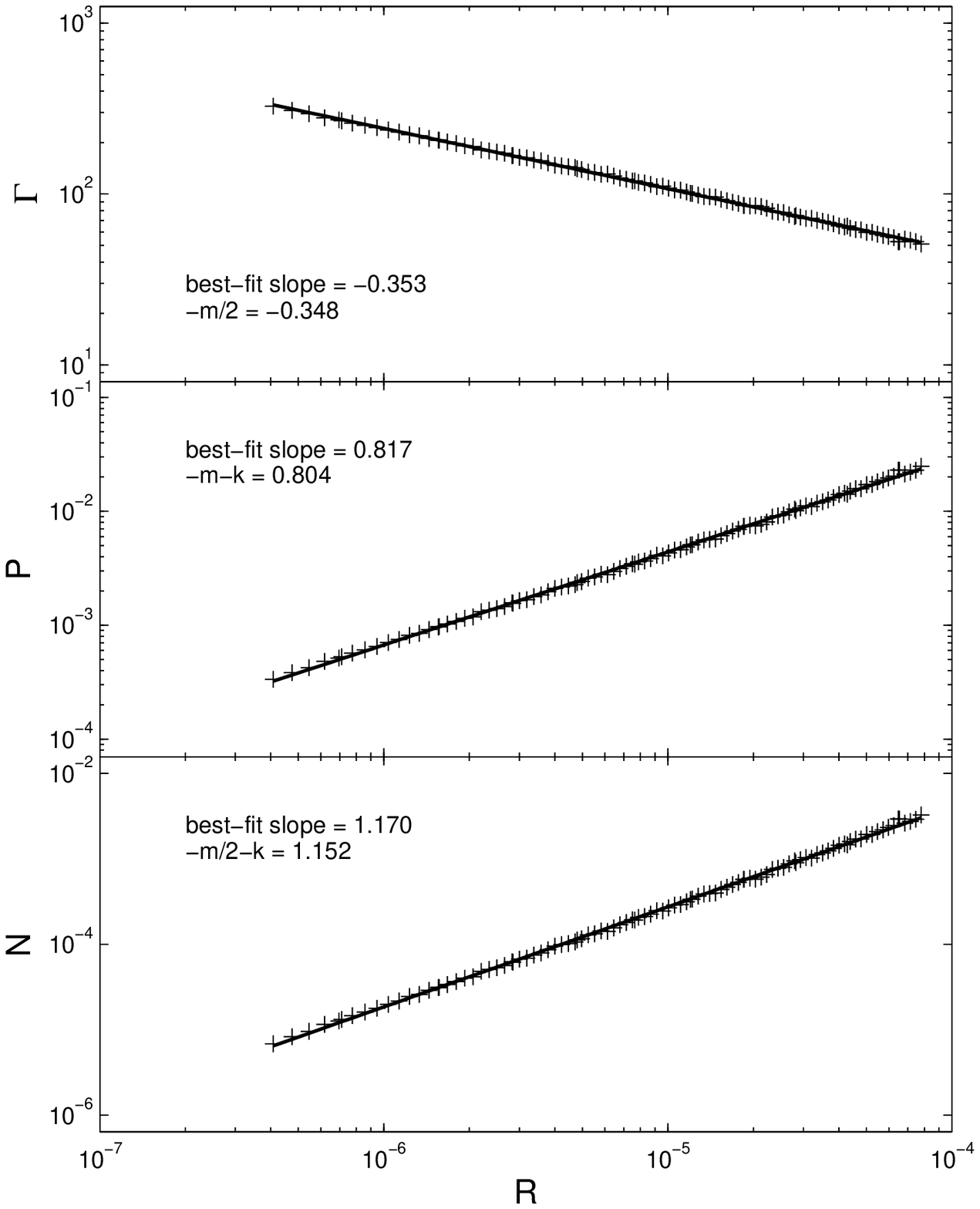}}}
\caption{Same as Figure~\ref{gpnvstime_in} but for times
after the shock emerges from the star. $\Gamma$, $P$, $N$, and $R$ were
deduced from the numerical data by finding at each time the position
where $\gamma^{-1}p/n$, the ratio of the thermal and the bulk kinetic
energies in the frame of the fluid, fell just below the constant
value we expect at the time of breakout.}
\label{gpnvstime_out}
\end{figure}

To verify our results numerically, we integrated the time-dependent
relativistic hydrodynamic equations using a one-dimensional
code. Figure~\ref{gprofile_in} shows curves for $\gamma$ as a function
of position at a single time before breakout, while
Figure~\ref{gpnvstime_in} shows the time evolution of $\Gamma$, $P$,
and $N$ before breakout. The numerical and analytic results are in
excellent agreement. Figures~\ref{gprofile_out} and
\ref{gpnvstime_out} respectively show the $\gamma$ vs. $x$ profile and
time evolution of $\Gamma$, $P$, and $N$ after breakout; the agreement
between numerical and analytic results here confirms the choice of
scale $R(t)$ after breakout that we discussed in \S~4.1.

\section{Summary}

We have shown that, given an ultrarelativistic shock propagating into
a planar polytropic envelope, the flow upon the shock's emergence from
the envelope into vacuum follows a self-similar solution strikingly
similar to the self-similar solution describing the flow while the
shock remains within the envelope. Both self-similar solutions obey
the same relations with regard to the time-evolution of the
characteristic quantities $R$, $\Gamma$, $P$, $N$ and with regard to
the similarity variables $\chi$, $g$, $f$, $h$. The pre- and
post-breakout solutions differ only in that the applicable ranges in
$\chi$ and the physical interpretations of the characteristic
quantities differ. As a result of these differences, the behavior of
the flow after breakout lies somewhere between the traditional Type I
and Type II classes of self-similar solutions; before breakout a Type
II solution applies. To arrive at these results we have looked in
detail at the behavior when the shock reaches the outer edge of the
envelope.

We have discussed these results in the context of an application---the
motion of a shock wave through a polytropic envelope near the surface
of a star, the shock's emergence from the surface, and the subsequent
flow into vacuum. This situation may be related to the explosions
believed to cause gamma-ray bursts and supernovae \cite[see, for
example,][]{tan01} and should be especially relevant in very optically
thick media such as neutron stars.

\acknowledgements
This research was partially funded by a NASA ATP grant.
RS is a Packard Fellow and an Alfred P. Sloan Research Fellow.


\begin{thebibliography}{11}
\expandafter\ifx\csname natexlab\endcsname\relax\def\natexlab#1{#1}\fi

\bibitem[{{Best} \& {Sari}(2000)}]{best00}
{Best}, P., \& {Sari}, R. 2000, Physics of Fluids, 12, 3029

\bibitem[{{Blandford} \& {McKee}(1976)}]{blandford76}
{Blandford}, R.~D., \& {McKee}, C.~F. 1976, Physics of Fluids, 19, 1130

\bibitem[{{Johnson} \& {McKee}(1971)}]{johnson71}
{Johnson}, M.~H., \& {McKee}, C.~F. 1971, \prd, 3, 858

\bibitem[{{Nakayama} \& Shigeyama(2005)}]{nakayama05}
{Nakayama}, K., \& Shigeyama, T. 2005, astro-ph 0503252

\bibitem[{{Sari}(2005)}]{sari05}
{Sari}, R. 2005, submitted

\bibitem[{{Sedov}(1946)}]{sedov46}
{Sedov}, L.~I. 1946, Appl. Math. Mech. Leningrad, 10, 241

\bibitem[{{Tan} {et~al.}(2001){Tan}, {Matzner}, \& {McKee}}]{tan01}
{Tan}, J.~C., {Matzner}, C.~D., \& {McKee}, C.~F. 2001, \apj, 551, 946

\bibitem[{{Taylor}(1950)}]{taylor50}
{Taylor}, G.~I. 1950, Proc. R. Soc. London Ser. A, 201, 175

\bibitem[{{von Neumann}(1947)}]{vonneumann47}
{von Neumann}, J. 1947, Blast Waves, Tech. Rep.~7, {Los Alamos National
  Laboratories}

\bibitem[{{Waxman} \& {Shvarts}(1993)}]{waxman93}
{Waxman}, E., \& {Shvarts}, D. 1993, Physics of Fluids, 5, 1035

\bibitem[{{Zel'dovich} \& {Raizer}(1967)}]{zeldovich67}
{Zel'dovich}, Y.~B., \& {Raizer}, Y.~P. 1967, {Physics of Shock Waves and
  High-Temperature Phenomena} ({New York}: {Academic Press})

\end{thebibliography}

\newcommand{\noopsort}[1]{} \newcommand{\printfirst}[2]{#1}
  \newcommand{\singleletter}[1]{#1} \newcommand{\switchargs}[2]{#2#1}

\end{document}